\documentstyle[floats,aps,pre,psfig]{revtex}
\newcommand{\be}{\begin{equation}}
\newcommand{\ee}{\end{equation}}
\newcommand{\bea}{\begin{eqnarray}}
\newcommand{\eea}{\end{eqnarray}}
\newcommand{\les}{\ell_{\hbox{\tiny ES}}}

\newcommand{\peff}{p_n^{\hbox{\tiny eff}}}

\newcommand{\ps}{p^{\hbox{\tiny S}}}
\newcommand{\pu}{p^{\hbox{\tiny U}}}
\newcommand{\Pdep}{P_{\hbox{\tiny dep}}}

\newcommand{\oni}{\omega_{\hbox{\tiny NI}}}
\newcommand{\omf}{\omega_{\hbox{\tiny MF}}}

\newcommand{\rhoC}{\rho_{\hbox{\tiny C}}}
\newcommand{\ttr}{\tau_{tr}}
\newcommand{\tdep}{\tau_{dep}}
\newcommand{\tres}{\tau_{res}}
\newcommand{\tdis}{\tau_{dis}}

\newcommand{\Wni}{W_{\hbox{\tiny NI}}}
\newcommand{\Pni}{P_{\hbox{\tiny NI}}}
\newcommand{\Qni}{Q_{\hbox{\tiny NI}}}
\newcommand{\fra}[2]{\hbox{${#1\over #2}$}}
\newcommand{\bold}[1]{\hbox{\bf x}}

\begin{document}
\draft

\title{The process of irreversible nucleation in multilayer growth. \\
II. Exact results in one and two dimensions.}

\author{Paolo Politi$^{1,\dag}$ and Claudio Castellano$^{2,*}$}
\address{$^1$ Istituto Nazionale per la Fisica della Materia, Unit\`a di
Firenze, Via. G. Sansone 1, 50019 Sesto Fiorentino, Italy}
\address{$^2$ Istituto Nazionale per la Fisica della Materia, 
Unit\`a di Roma 1\\ and
Dipartimento di Fisica, Universit\`a di Roma
``La Sapienza'', Piazzale Aldo Moro 2, 00185 Roma, Italy}

\author{(\today)}
\author{\parbox{397pt}{\vglue 0.3cm \small
We study irreversible dimer nucleation on top of 
terraces during epitaxial growth in one and two dimensions,
for all values of the step-edge barrier.
The problem is solved exactly by transforming it into a first
passage problem for a random walker in a higher-dimensional space.
The spatial distribution of nucleation events is shown to differ
markedly from the mean-field estimate except in the limit of 
very weak step-edge barriers.
The nucleation rate is computed exactly, including numerical prefactors.
}}
\maketitle


\section{Introduction}
\label{Intro}

The understanding of how atomistic processes influence morphology at
large scales is of fundamental importance for controlled
growth of crystalline films via deposition techniques.
The irreversible nucleation of immobile dimers, giving rise to new
terraces, is a key process for the growth of a high symmetry surface.
In the preceding paper\cite{Politi01} we have shown that for the nucleation
on top of existing terraces, the usual Mean Field (MF)
theory~\cite{mft,Tersoff94} is equivalent to considering particles as
noninteracting: i. e. not feeling each other even if they
are on the same lattice site, so that they can meet several times
before leaving the terrace. Mean Field Theory (MFT) 
counts all these fictitious nucleation
events and therefore leads to an overestimate of the nucleation
rate $\omega$, that in most cases is a very poor
approximation of the correct results.
For the spatial distribution of nucleation events we have shown in
Ref.~\cite{Politi01}
that a substantial discrepancy between mean-field and exact results
is expected, because fictitious nucleations beyond the first one
always dominate in $d=1$ and $d=2$.

In this paper we go beyond mean-field theory and present a series
of exact results.
We calculate the spatial [$P(n)$] and temporal [$Q(t)$]
distributions of nucleation events. 
The quantity $Q(t)$
is the probability that two atoms meet a time $t$ after deposition
of the second atom and it is
formally defined in Eq.~(\ref{def_PQ}).
The evaluation of $P$ and $Q$ allows the determination of the 
total probability $W$ that two atoms meet and this allows
the exact computation of the nucleation rate $\omega$.

The solution of the problem
is obtained by mapping the diffusion of two particles
on a $d$-dimensional terrace into the motion of a single random walker
in $d'=2d$ dimensions.
The statistics of meeting events between the two adatoms (nucleations)
is then obtained as the solution of a suitable first passage problem for
the $d'$-dimensional random walker.
In $d=1$ the problem can be treated analytically
in full detail, leading to closed form expressions for all quantities
of interest.
In $d=2$ one can easily obtain the results numerically, with arbitrary
accuracy.

Results indicate that the spatial distribution of nucleation
sites is very different from the mean-field estimate in the limit
of strong Ehrlich-Schwoebel barriers, both in $d=1$ and $d=2$.
In the opposite limit of zero or weak barriers instead,
the difference between the mean-field estimate and the exact result
is, for reasonable terrace sizes, quite surprisingly small.
The temporal distribution of nucleation events decays slowly
for short times and later exponentially. 
Finally, the calculation of the nucleation rate $\omega$ 
is completed by the rigorous determination of the nucleation
probability $W=\sum_n P(n)$. This confirms that MFT is
safely applicable only for weak barriers in $d=2$ and gives
the exact expressions for $\omega$ that must be used instead
of the MF approximate ones.

The paper is organized as follows. In Sec.~\ref{method} the model
for irreversible nucleation is presented and the fundamental
quantities needed in the rest of the paper are introduced.
The method for the solution of the problem is also outlined.
Secs.~\ref{d1} and~\ref{d2} are devoted to the presentation
of the exact results obtained in $d=1$ and $d=2$, respectively.
In Sec.~\ref{disc} these results are discussed and interpreted in
physically intuitive terms.
The conclusions and the perspectives of this work can be
found in Sec.~\ref{concl}.

Some of the most important results have been presented
previously in Ref.~\cite{Castellano01}.

\section{The problem and the method of solution}
\label{method}

In this Section we briefly recall the basic concepts of
irreversible dimer nucleation along with some results, obtained
in the first paper, that will be needed in the following.

We consider particles deposited onto a crystalline terrace of size
$L$, modeled as a discrete lattice (a square lattice in $d=2$).
The flux of particles is uncorrelated, uniform and of
intensity $F$, so that the average interarrival time is
$\tdep=(FL^d)^{-1}$.
Once on the terrace, an adatom hops at rate $(\Delta t)^{-1} = 2dD$ 
to a randomly chosen nearest neighbor, until it either meets another
adatom or leaves the terrace.

This last process can be hindered by the additional Ehrlich-Schwoebel
(ES) barrier~\cite{seb} reducing interlayer transport to a rate $2dD'$:
the ES length $\les=(\fra{D}{D'}-1)a_0$ measures the strength of the 
barrier (in the following the lattice constant $a_0$ is used as unit length).

The average time spent by a single adatom on the terrace is the residence
time and depends on $L$ and $\les$,
\be
\tres = (\beta L+\alpha\les) L/D ~.
\label{eq_tres}
\ee
In the limit $\les=0$, $\tres$ is equal to $\ttr$, the average time needed
by an adatom to reach the terrace boundary.
Depending on the value of $\les$, three different regimes may occur:
i) Zero or weak barriers ($\ttr \simeq \tres \ll \tdep$);
ii) Strong barriers ($\ttr \ll \tres \ll \tdep$);
iii) Infinite barriers ($\ttr \ll \tdep \ll \tres$).

Particles are deposited according to an exponential distribution
of interarrival times: $\Pdep(\tau)=\tdep^{-1}\exp(-\tau/\tdep)$.
This implies that all quantities should be computed for a generic 
interarrival time $\tau$
and then the results should be averaged over $\Pdep(\tau)$.
However, we have shown in Ref.~\cite{Politi01} that this is equivalent
to considering two particles deposited simultaneously, one with distribution
$\pu_n=1/L^d $ and the other with an effective distribution
\be
\peff ={\tres \over \tdep + \tres} \ps_n ~ ,
\ee
where $\ps_n$ is the normalized solution of the discrete stationary
diffusion equation in the presence of a constant flux.
For infinite barriers (regime {\em iii}) $\peff=\ps_n=1/L^d$.
For strong but finite barriers (regime {\em ii})
$\peff =  {\tres \over \tdep} \ps_n =
{\tres \over \tdep} {1 \over L^d}$.
In the limit of zero or weak barriers (regime {\em i}) $\peff =
{\tres \over \tdep} \ps_n$ where $\ps_n$ has a parabolic shape that
vanishes at the edges, reflecting the presence of absorbing boundaries.

Nucleation of dimers takes place when
particles are on adjacent lattice sites; here we will assume instead that
a dimer is formed when two particles are on the same site:
this avoids useless mathematical complications without modifying
the physics of the nucleation process.

The physical quantities we are interested in are $P(n)$, $Q(t)$ and
$\omega$.

$\bullet$ $P(n)$ is the spatial distribution of nucleation events, computed for
two adatoms deposited at the same time with normalized distributions
$\ps$ (the first) and $\pu$ (the second).
$P^{(N)}(n)$ is its normalized version.

$\bullet$ The distribution $Q(t)$ is the probability that a nucleation 
event occurs at time $t$, if the two adatoms have been
deposited at time zero. 
$Q(t)$ is not considered within the standard mean-field theory.

$P(n)$ and $Q(t)$ are derived from the same quantity, the probability
$R(n,t)$ that a nucleation event occurs on site $n$ at time $t$:
\be
P(n) = \sum_t R(n,t) ~ ,
\hspace{2cm}
Q(t) = \sum_n R(n,t) ~ .
\label{def_PQ}
\ee

We can also define $W$, the probability that two atoms meet before leaving
the terrace: it is clearly related to $P(n)$ and $Q(t)$, because
\be
W \equiv \sum_{n,t} R(n,t) = \sum_n P(n) = \sum_t Q(t) ~ .
\ee

$W$ is equal to 1 for large or infinite barriers
(regimes {\it ii} and {\it iii}), but it differs from unity in regime
{\it i}.
The normalized spatial distribution is clearly $P^{(N)}(n)=P(n)/W$.

$\bullet$ The nucleation rate $\omega$ is the total number of nucleation events
that occur on the whole terrace per unit time.
It is related to $P(n)$ or $Q(t)$ via $W$:
\be
\omega = F L^d {\tres \over \tdep + \tres} W~.
\label{omegagen}
\ee

$P(n)$ and $Q(t)$ (and then $W$) depend on the {\em normalized} initial
distributions for the two adatoms.
Therefore, they have the same expressions in regime {\em ii} and {\em iii},
where $\ps_n$ is just a constant. From
Eq.~(\ref{omegagen}) instead, one immediately realizes that
$\omega$ has different expressions in each of the three regimes.

We also consider an artificial model
where adatoms are independent diffusing particles. They do not stop
when they meet and each encounter is considered as a (fictitious) nucleation.
As shown in Ref.~\cite{Politi01}, this model gives exactly the same
results as mean-field theory.

The computation of the quantities of interest requires the evaluation of
$R(n,t)$.
Since we consider irreversible dimer formation, $R(n,t)$ is the probability
that two particles diffusing on a $d$-dimensional terrace
meet for the first time on site $n$ at time $t$.
A method for treating the diffusion of two particles
is to take their $d+d$ coordinates as the coordinates of a
single random walker diffusing on a $d'=2d$-dimensional hypercubic
terrace.
In this picture a nucleation event corresponds to
the $d'-$dimensional walker reaching the $d$-dimensional hyperplane
where the coordinates of the two particles are equal.
The irreversibility of dimer formation implies that an absorbing
boundary condition must be imposed on this $d-$dimensional hyperplane.
The probability of dimer formation $R(n,t)$ is then given by the probability
current orthogonal to the hyperplane.

More specifically, in $d=1$ we pass from two walkers of coordinates
$n$ and $m$ to a single walker on a square terrace. Nucleation occurs
when the walker reaches the diagonal of such a terrace ($n = m$).
In $d=2$ we must consider a single walker in a four-dimensional space whose
coordinates are $(n_1,m_1,n_2,m_2)$ and the hyperplane is now a
bidimensional plane defined by the conditions $n_1 = n_2$ and $m_1=m_2$.

In this way we have reduced the dimer nucleation problem to a first
passage problem~\cite{Redner}.
The solution of such a problem~\cite{origin} 
is possible analytically in $d=1$ 
(Sec.~\ref{d1}) and numerically in higher dimension (Sec.~\ref{d2}).

\section{Results in one dimension}
\label{d1}

When the system is one-dimensional, the two adatoms are mapped into a 
two-dimensional walker hopping inside a square lattice
of size $L$ with probability $p_{m,n}(t)$ to be in
site $(m,n)$ at time $t$. 
Assuming that, when two adatoms are present, one of them, randomly chosen,
moves once every time unit, 
the discrete evolution equation for $p_{m,n}(t)$ is
\be
p_{m,n}(t+1)={1\over 4}[p_{m+1,n}(t)+p_{m-1,n}(t)+p_{m,n+1}(t)
+p_{m,n-1}(t)]
\label{eq2d}
\ee
where the discrete time unit corresponds now to a physical time 
$\Delta t = 1/(2d'D) = 1/(4 d D)$.

The indices $m,n$ vary between 1 and $L$, but in order to use
Eq.~(\ref{eq2d}) for all terrace sites it is useful to introduce
fictitious sites in $m=0,L+1$ and $n=0,L+1$.
In this way, boundary conditions are easily written for
generic values of the ES barrier:
$p_{0,n}=a p_{1,n}$, $p_{L+1,n}=a p_{L,n}$,
$p_{m,0}=a p_{m,1}$, $p_{m,L+1}=a p_{m,L}$, where
$a=\les/(1+\les)$.
They apply at any time and for all edge sites.
There is also an additional boundary condition along the square
diagonal
\be
p_{n,n}(t)=0 ~~~~~~ n=1,\dots,L ~ ,
\ee
because the two adatoms stop diffusing when they meet.
The initial condition is $p_{m,n}(0)= \pu_m \ps_n$, but it is also
correct to write $p_{m,n}(0)= \ps_m \pu_n$. 
In order to obtain a spatial distribution $P(n)$ that is properly 
symmetrical with respect to the center of the terrace we use the
symmetrized expression:
\be
p_{m,n}(0)= \fra{1}{2} [\pu_m \ps_n + \ps_m \pu_n] ~ .
\label{pmn0}
\ee

The basic quantity we want to compute, the nucleation probability on
site $n$ at time $t+1$, is
\be
R(n,t+1)={1\over 4}[p_{n+1,n}(t)+p_{n-1,n}(t)+p_{n,n+1}(t) +p_{n,n-1}(t)] ~ . 
\label{Rint}
\ee

In the case of noninteracting particles, the boundary condition along the
diagonal is dropped and Eq.~(\ref{Rint}) is replaced by
\be
R(n,t)=p_{n,n}(t) ~ .
\ee

An explicit analytic solution of the problem,
both for interacting and noninteracting particles, is possible
in the limits of zero and infinite ES barriers and will be presented
in detail below.
As remarked in Sec.~\ref{method}, for $P$, $Q$ and $W$
only two distinct regimes exist, and $\les=0,\infty$ are their
representative limits.

\subsection{Zero barriers (regime {\em i})}

When no ES barrier is present, $\les=0$ and $a=0$. Hence the boundary
conditions are simply
$p_{0,n}=p_{L+1,n}=p_{m,0}=p_{m,L+1}=0$,
indicating that edges are absorbing boundaries.
In the limit $\les=0$ the normalized stationary distribution is
\be
\ps_n = {6 \over L(L+1) (L+2)} n(L+1-n) ~ .
\ee

\subsubsection{Noninteracting adatoms}

By separating space and time variables in a way perfectly analogous
to the treatment of a single particle~\cite{Politi01},
we find the general solution of Eq.~(\ref{eq2d})
\be
p_{m,n}(t) = \sum_{k,j=1}^L B_{kj}{1\over 2^t}\left[
\cos\left({k\pi\over L+1}\right) + \cos\left({j\pi\over L+1}\right)\right]^t
\sin\left({m k\pi\over L+1}\right)
\sin\left({n j\pi\over L+1}\right)
\label{sol2d}
\ee
where the coefficients $B_{kj}$ are
\be
B_{kj} = \left({2\over L+1}\right)^2 \sum_{m,n=1}^L p_{m,n}(0)
\sin\left({m k\pi\over L+1}\right)
\sin\left({n j\pi\over L+1}\right).
\label{coeff}
\ee

Given the explicit form~(\ref{pmn0}) of $p_{m,n}(0)$, the coefficients
$B_{kj}$ are (see Ref.~\cite{Politi01})
\bea
B_{kj} & = & A_k^U A_j^S \\
& = & {12 \over L^2 (L+1)^3 (L+2)} 
{\sin\left({k \pi \over 2}\right) \sin\left({j \pi \over 2}\right) 
\over 
\sin\left[{k \pi \over 2(L+1)}\right]
\sin^3\left[{j \pi \over 2(L+1)}\right]}
\sin\left[{L k \pi \over 2(L+1)}\right]
\sin\left[{L j \pi \over 2(L+1)}\right]
\eea
and this allows the evaluation of all the quantities of interest.

\begin{figure}
\centerline{\psfig{figure=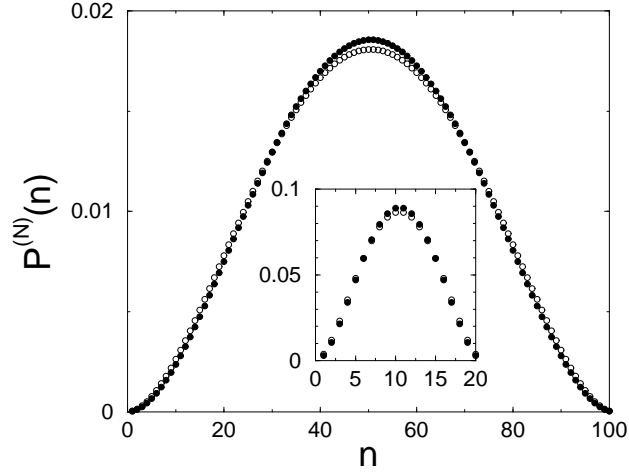,width=8cm,angle=-90}}
\caption{Normalized spatial distribution $P^{(N)}(n)$ for $d=1$ and
$\les=0$. Empty circles are for interacting particles, full circles
for noninteracting particles (Mean Field Theory). 
$L=100$ in the main part, $L=20$ in the inset.}
\label{fig1}
\end{figure}

\noindent $\bullet$ {\it Nucleation sites.}
The spatial distribution of nucleation sites is:
\be
\Pni(n) = \sum_{t=0}^\infty R(n,t) = \sum_{t=0}^\infty p_{n,n}(t) ~ . 
\ee
Its normalized version $\Pni^{(N)}(n)$ is plotted in Fig.~\ref{fig1}.
As proven in Ref.~\cite{Politi01} within a continuum formalism,
it is equal to the mean field distribution.
This result can be easily proven analytically in a discrete lattice as well.

\begin{figure}
\centerline{\psfig{figure=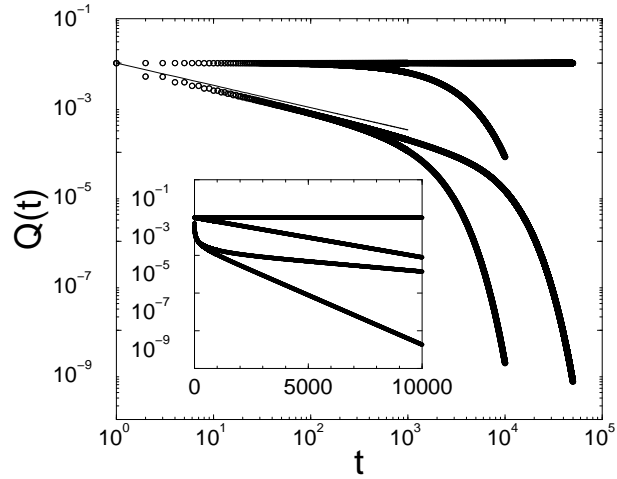,width=8cm,angle=-90}}
\caption{The temporal distribution $Q(t)$ for $d=1$ and $L=100$.
From top to bottom, data are for noninteracting particles ($\les=\infty$
and $\les=0$) and interacting particles ($\les=\infty$ and $\les=0$).
The main part of the figure highlights the power-law decay for short 
times (log-log plot): the solid line goes as $t^{-1/2}$.
The inset highlights the exponential decay for long times (lin-log plot). }
\label{fig2}
\end{figure}

\noindent $\bullet$ {\it Nucleation times.}
The distribution of nucleation times is given by:
\begin{eqnarray}
\Qni(t) & = & \sum_{n=1}^L p_{n,n}(t) \\
& = & \sum_{k,j=1}^L B_{kj} \left\{{1 \over 2}
\left[\cos\left({k \pi \over L+1}\right)
+\cos\left({k \pi \over L+1}\right) \right] \right\}^t
\sum_{n=1}^L
\sin\left({n k \pi\over L+1}\right)
\sin\left({n j \pi\over L+1}\right) \\
& = & {L+1 \over 2} \sum_{k=1}^L B_{kk}
\cos^t\left({k \pi \over L+1}\right)
\end{eqnarray}
and it is plotted in Fig.~\ref{fig2}.
To find analytically the behavior of $\Qni(t)$ for large $L$,
we rewrite it in the following form:
\be
\Qni(t) = {L+1 \over 2} \sum_{k=1}^L B_{kk} \exp
\left[t \ln \cos\left({k \pi \over L+1} \right) \right] ~ .
\ee

The coefficients $B_{kk}$ diverge for small $k$
as $k^{-4}$ ($B_{kk} \simeq 192/[\pi^4 k^4 (L+1)(L+2) L^2]$), so that
the dominant contribution to the sum for large $L$
comes from the first mode $k=1$. Expanding the small argument of the
cosine this gives
\be
\Qni(t) \sim \exp \left[-{t \over 2} \left({\pi \over L+1} \right)^2 \right]
 ~ ,
\label{qni1db0}
\ee
which is exactly the exponential decay appearing in Fig.~\ref{fig2}.

\noindent $\bullet$ {\it Nucleation rate.}
Let us first compute $\Wni=\sum_n P(n)= \sum_t Q(t)$ that for
noninteracting particles is not a probability but the total
number of times the two adatoms meet before leaving the terrace
($\Wni$ can be larger than 1):
\be
\Wni = {L+1 \over 2} \sum_{k=1}^L
{B_{kk} \over 1-\cos\left({k \pi \over L+1} \right)} ~ .
\label{WNI}
\ee
Using the explicit form of $B_{kk}$, we obtain
\be
\Wni = {3 \over L^2 (L+1)^2 (L+2)} \sum_{k=1}^L
{\sin^2\left({k \pi \over 2}\right)
\sin^2\left({L k \pi \over 2(L+1)}\right)
\over 
\sin^6\left({k \pi \over 2(L+1)}\right)} ~ .
\ee
For large $L$ the dominant contribution is provided by
the term with $k=1$, which gives
\be
\Wni \simeq {3 \over L^2 (L+1)^2 (L+2)}
\left[{2 (L+1) \over \pi}\right]^6 = 3 \left({2 \over \pi}\right)^6
{(L+1)^4 \over L^2 (L+2)} \simeq 0.2 L ~ .
\label{Wnid1}
\ee
Hence the total nucleation rate is
\be
\oni \simeq F L {\tau_{res} \over \tau_{dep}}
3 \left({2 \over \pi}\right)^6 {(L+1)^4 \over L^2 (L+2)} ~ .
\ee
Using the explicit expression~\cite{Politi01} $\tres=\ttr=L^2/(12 D)$,
and considering only the leading order in $L$ we find
\be
\oni \simeq {1 \over 4} \left({2 \over \pi}\right)^6 {F^2 L^5 \over D} ~ .
\ee

\subsubsection{Interacting adatoms}

For interacting adatoms 
it is possible to take advantage of the noninteracting
solution (\ref{sol2d}) by using a trick: we pass from the initial
condition $p_{m,n}(0)$ [given in Eq.~(\ref{pmn0})]
to an auxiliary antisymmetric initial condition
\be
{\tilde p}_{m,n}(0) = \left\{ \begin{array}{rcl}
p_{m,n}(0) & \mbox{~~for~~}m<n \\
0 & \mbox{~~for~~}m=n \\
-p_{n,m}(0) & \mbox{~~for~~}m>n
\end{array}
\right.
\ee
which satisfies the boundary condition ${\tilde p}_{n,n}=0$
along the diagonal.

Let us observe that the dynamics given in Eq.~(\ref{eq2d})
conserves the parity of the spatial distribution, because:
\be
{\tilde p}_{m,n}(0)=- {\tilde p}_{n,m}(0) \Rightarrow B_{kj}=- B_{jk}
\Rightarrow {\tilde p}_{m,n}(t)=- {\tilde p}_{n,m}(t) ~ .
\ee

This means that if we start with an antisymmetric distribution
${\tilde p}_{m,n}(0)=-{\tilde p}_{n,m}(0)$, 
the boundary condition ${\tilde p}_{n,n}(t)=0$
is obeyed for all times: the two triangles
$(m>n)$ and $(m<n)$ are dynamically disconnected.
The solution of Eq.~(\ref{eq2d}) is therefore
still given by Eq.~(\ref{sol2d}), since 
the boundary condition is fully taken into account by the
value of the coefficients $B_{kj}$, which depends on the antisymmetric
form of ${\tilde p}_{m,n}(0)$.

The coefficients $B_{kj}$ are given by
\be
B_{kj} = \left({2\over L+1}\right)^2 \sum_{m,n=1}^L {\tilde p}_{m,n}(0)
\sin\left({m k\pi\over L+1}\right)
\sin\left({n j\pi\over L+1}\right).
\ee
In this expression we can decompose the summation $\sum_{m,n}$
as $\sum_{m<n} +
\sum_{m>n}$, in the latter interchange the dumb indices $n,m$ 
and exploit the antisymmetry of ${\tilde p}_{m,n}(0)$.
We finally obtain:
\be
B_{kj} = \left({2\over L+1}\right)^2
\sum_{m<n}p_{m,n}(0)\left\{
\sin\left({m k\pi\over L+1}\right)
\sin\left({n j\pi\over L+1}\right) - (k \Leftrightarrow j) \right\} 
\equiv  [ B_{kj}^< - B_{jk}^< ]
\ee
where
\be
B_{kj}^< =  \left({2\over L+1}\right)^2
\sum_{m<n}p_{m,n}(0) \sin \left({m k\pi\over L+1}\right)
\sin \left({n j\pi\over L+1}\right) ~ .
\ee

The evaluation of $B_{kj}$ is here less straightforward than in the
noninteracting case. In particular, some sums are not easily performed
explicitly. This makes difficult the presentation of explicit results.
Therefore in the following we will present only the general results,
leaving the coefficients $B_{kj}$ indicated.

\noindent $\bullet$ {\it Nucleation sites.}
Since the regions $m<n$ and $m>n$ are equivalent,
the probability of a nucleation event on site $n$ at time $t+1>0$ is given
by $R(n,t+1)={1 \over 2} [{\tilde p}_{n,n+1}(t)+{\tilde p}_{n-1,n}(t)]$,
while for $t=0$ nucleations occur because both adatoms are deposited on
the same site, i. e. with probability $p_{n,n}(0)$.
The spatial distribution of nucleation sites is therefore
\begin{eqnarray}
P(n) & = & p_{n,n}(0) +  {1 \over 2} \sum_{t=0}^\infty
[{\tilde p}_{n,n+1}(t) + {\tilde p}_{n-1,n}(t)] \\
& = & p_{n,n}(0)  + {1 \over 2}
\sum_{k,j=1}^L {B_{kj} \over 1-{1 \over 2}
\left[\cos\left({k \pi \over L+1}\right)
+ \cos\left({j \pi \over L+1} \right) \right]} \\
& & \left\{
\sin\left({n k \pi \over L+1}\right) \sin\left[{(n+1) j \pi \over L+1}\right] +
\sin\left[{(n-1) k \pi \over L+1}\right] \sin\left({n j \pi \over L+1}\right)
\right\}
\end{eqnarray}
and the normalized distribution $P^{(N)}(n)$ is presented in Fig.~\ref{fig1}.
The plot clearly shows that the spatial distribution
is very similar to the mean-field result $\Pni^{(N)}(n)$,
although a small discrepancy exists
(see the discussion in Sec.~\ref{disc_Pdin}).

\noindent $\bullet$ {\it Nucleation times.}
The distribution of nucleation times for interacting particles is
\begin{eqnarray}
Q(0) & = &  \sum_{n=1}^L p_{n,n}(0) ~ ,\\
Q(t+1>0) & = & \sum_{n=1}^L {1 \over 2} 
[{\tilde p}_{n,n+1}(t) + {\tilde p}_{n-1,n}(t)] ~ .
\label{Qtp1}
\end{eqnarray}
Once summed, the two terms in Eq.~(\ref{Qtp1}) are equal. Hence,
\begin{eqnarray}
Q(t+1>0) & = & \sum_{n=1}^L {\tilde p}_{n-1,n}(t) \\
& = & \sum_{k,j=1}^L B_{kj} C_{kj} \left\{ {1 \over 2} \left[
\cos\left({k \pi \over L+1} \right) +
\cos\left({j \pi \over L+1} \right) \right] \right\}^t ~ ,
\end{eqnarray}
where the coefficients $C_{kj}$ are
\begin{eqnarray}
\label{C}
C_{kj} & = & \sum_{n=1}^L 
\sin\left[{k \pi (n-1)\over L+1} \right]
\sin\left({j \pi n\over L+1} \right) \\
& = & {1 \over 2} \left\{
\cos\left[{k \pi \over L+1} + {L (j-k) \pi \over 2(L+1)}\right]
\sin\left[{(j-k) \pi \over 2}\right] \csc\left[{(j-k)\pi \over 2(L+1)}\right]
\right .\\
& - &
\left . \cos\left[-{k \pi \over L+1} + {L (j+k) \pi \over 2(L+1)}\right]
\sin\left[{(j+k) \pi \over 2}\right] \csc\left[{(j+k) \pi \over 2(L+1)}\right]
\right\} ~ .
\end{eqnarray}

$Q(t)$ is shown in Fig.~\ref{fig2}.
For short times it decays slowly (as $t^{-1/2}$), while it goes down
exponentially for large times.
In Appendix~\ref{App2} we show in detail that the behavior of $Q(t)$ for short
and long times can be derived explicitly in the case of two adatoms
with uniform initial distributions, for which the coefficients
$B_{kj}$ are explicitly known: for $t \ll 2 L^2/\pi^2$ we find
\be
Q(t)\simeq {8 \over L \pi^{5/2} \sqrt{t/2}}
\ee
and for $t\gg 2 L^2/\pi^2$
\be
Q(t)\simeq {80 \over 9 \pi^2 L^2} \exp
\left[-{5 t \over 4} \left({\pi \over L+1} \right)^2 \right] ~ .
\label{Qexp}
\ee
No qualitative change is expected if one atom is initially
distributed according to $\ps_n$ rather than $\pu_n$:
only prefactors are expected to be different and this is confirmed
by the behavior shown in Fig.~\ref{fig2}.

\noindent $\bullet$ {\it Nucleation rate.}
The probability $W$ of a nucleation event is
\be
W = {1 \over L} + \sum_{k,j=1}^L {B_{kj} C_{kj} \over
1-{1 \over 2} \left[\cos\left({k \pi \over L+1}\right)
+ \cos\left({j \pi \over L+1} \right) \right]} ~ .
\label{Wi}
\ee
In Appendix~\ref{App2} we prove that for large $L$, $W$ goes to a constant.
This constant is found numerically to be roughly equal to $0.47$.
Hence, for large $L$, the nucleation rate is
\be
\omega = F L {\beta L^2 \over D} F L W \simeq 0.04 {F^2 L^4 \over D} ~ .
\label{omega1d}
\ee

\subsection{Strong and infinite barriers (regimes {\em ii} and {\em iii})}
With infinite ES barriers, $\les=\infty$ and $a=1$.
Step edges are perfectly reflecting barriers and
boundary conditions are:
$p_{1,n}=p_{0,n}$, $p_{L,n}=p_{L+1,n}$,
$p_{m,1}=p_{m,0}$, $p_{m,L}=p_{m,L+1}$.
The normalized stationary distribution is simply $\ps_n = \pu_n
= 1/L$, because the distribution of the first adatom is still flat when
the second arrives.

The general solution for a two-dimensional walker is now
\be
p_{m,n}(t) = \sum_{k,j=0}^{L-1} B_{kj}{1\over 2^t}\left[
\cos\left({k\pi\over L}\right) + \cos\left({j\pi\over L}\right)\right]^t
X_k(m) X_j(n) ~ ,
\ee
where
\be
X_k(n)=\tan\left({k \pi \over 2 L}\right) \sin \left({n k \pi \over L}\right)
+\cos\left({n k \pi \over L}\right)
\ee
and the coefficients are
\be
B_{kj} = {1 \over N_k N_j} \sum_{m,n=1}^{L} p_{m,n}(0) X_k(m) X_j(n)
\ee
with ($\delta_{k0}$ is the Kronecker symbol):
\be
N_k = {L \over 2}
\left[1 +\tan^2\left({k \pi \over 2 L}\right)  \right] ( 1+\delta_{k0}) ~ .
\ee

\subsubsection{Noninteracting adatoms}

The case with noninteracting adatoms is completely trivial for
infinite barriers. At any time, $p_{m,n}(t)=1/L^2$ so that
the spatial and temporal distributions of nucleation events are constant.
The total number of nucleation events $\Wni$ is clearly infinite.

\subsubsection{Interacting adatoms}

In a way analogous to the case with zero barriers,
we consider antisymmetric initial conditions and we obtain
\be
B_{kj} = {1\over N_k N_j} \sum_{m<n} p_{m,n}(0)
\left\{X_k(m) X_j(n) - (k \Leftrightarrow j) \right\} 
\equiv  [ B_{kj}^< - B_{jk}^< ]
\ee
where
\be
B_{kj}^< =  {1\over N_k N_j} \sum_{m<n} 
p_{m,n}(0) X_k(m) X_j(n) ~ .
\ee

More explicitly ($p_{m,n}(0)=1/L^2$):
\bea
B_{kj}^< & = & {1\over L^2 N_k N_j} 
\sum_{n=1}^L X_j(n) \sum_{m=1}^{n-1} X_k(m) \\
& = & {1\over L^2 N_k N_j} 
\sum_{n=1}^L \left[\tan\left({j \pi \over 2 L}\right)
\sin\left({n j\pi\over L}\right) +\cos\left({n j\pi\over L}\right) \right] \\
& & \sum_{m=1}^{n-1} \left[\tan\left({k \pi \over 2 L}\right)
\sin\left({m k\pi\over L}\right) +\cos\left({m k\pi\over L}\right) \right]
 ~ .
\eea

\begin{figure}
\centerline{\psfig{figure=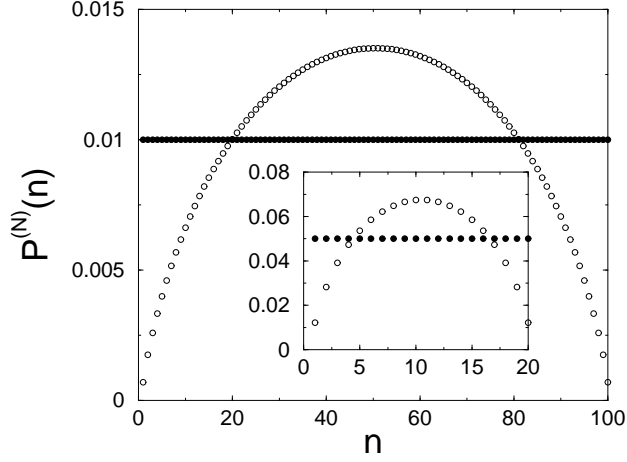,width=8cm,angle=-90}}
\caption{Normalized spatial distribution $P^{(N)}(n)$ for $d=1$ and
$\les=\infty$. Empty circles are for interacting particles, full circles
for noninteracting particles (mean-field approximation). 
$L=100$ in the main part, $L=20$ in the inset.}
\label{fig3}
\end{figure}

\noindent $\bullet$ {\it Nucleation sites.}
The distribution of nucleation sites is given by
\bea
P(n) & = & p_{n,n}(0) +  {1 \over 2} \sum_{t=0}^\infty
[{\tilde p}_{n,n+1}(t) + {\tilde p}_{n-1,n}(t)] \\
& = & {1 \over L^2}  + {1 \over 2}
\sum_{k,j=0}^{L-1} {B_{kj} \over 1-{1 \over 2}
\left[\cos\left({k \pi \over L+1}\right)
+ \cos\left({j \pi \over L+1} \right) \right]}
\left[X_k(n) X_j(n+1) + X_k(n-1) X_j(n) \right]
\label{Ples}
\eea
and it is plotted in Fig.~\ref{fig3}
[in this case $P^{(N)}(n)$ and $P(n)$ coincide, since $W=1$].
The distribution has a rounded
peak in the middle of the terrace and vanishes towards the boundaries.

The above expression for $P(n)$ is exact, but it is not easy to use
in applications.
A simpler, approximate, expression is therefore highly desirable.
In Sec.~\ref{disc_Pdin} we show that $P(n)$ is well fitted by a
hyperbolic cosine. Up to the normalization factor:
\be
P(n) = \cosh(\pi) -  \cosh \left[ \pi \left( {2 n \over L+1} - 1
\right) \right] ~ .
\label{Papprox}
\ee

\begin{figure}
\centerline{\psfig{figure=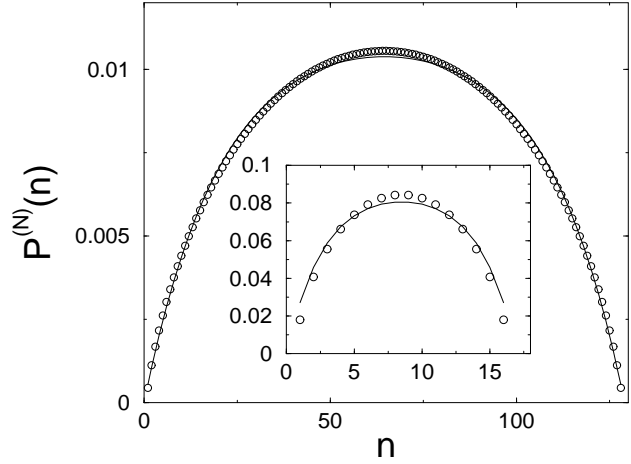,width=8cm,angle=-90}}
\caption{Comparison of the normalized spatial distribution
$P^{(N)}(n)$ for $d=1$ and $\les=\infty$ (circles) with the approximate
formula~(\ref{Papprox}) (solid line). $L=128$ (main), $L=16$ (inset).}
\label{fig4}
\end{figure}

In Fig.~\ref{fig4} we compare exact and approximate distributions:
the agreement is fair already for relatively small sizes and
very good for large sizes.

\noindent $\bullet$ {\it Nucleation times.}
As in the case with no barriers, we have
\bea
Q(0) & = &  \sum_{n=1}^L p_{n,n}(0) ~ , \\
Q(t+1>0) & = & \sum_{n=1}^L {\tilde p}_{n-1,n}(t) ~ .
\eea

Using the expression for ${\tilde p}_{m,n}(t)$
\bea
Q(0) & = &  {1 \over L} ~ , \\
Q(t+1>0) & = & \sum_{k,j=0}^{L-1} B_{kj} C_{kj}
\left\{ {1 \over 2} \left[
\cos\left({k \pi \over L} \right) +
\cos\left({j \pi \over L} \right) \right] \right\}^t
\eea
where the coefficients $C_{kj}$ are now
\be
C_{kj} =  \sum_{n=1}^L X_k(n-1) X_j(n)
\ee
The form of $Q(t)$ is shown in Fig.~\ref{fig2}.
The decay is the same as for zero barriers: for short times it decays
as $t^{-1/2}$ and for large times exponentially.
Physically intuitive interpretations of these behaviors 
are discussed in Sec.~\ref{disc_Qdit}.

\noindent $\bullet$ {\it Nucleation rate.}
Since $W=1$, the nucleation rate in regime {\em ii} is
\be
\omega(L) = F L {\tres \over \tdep} = {F^2 L^3 \les \over 2 D}
\label{omega1dles}
\ee
while in regime {\em iii} it is simply
\be
\omega(L) = F L = {1 \over \tdep} ~ .
\ee

\subsection{Intermediate barriers}
\label{intermediate}

For intermediate values of the barriers, i. e. values of 
$a$ between 0 and 1, an explicit analytic solution of the
problem is not possible, even for noninteracting adatoms.
This is a direct consequence of the lack of an explicit solution
for intermediate barriers even in the case of a single particle
(Ref.~\cite{Politi01}).
Nevertheless the problem can easily be solved numerically for any $\les$,
through direct calculation of the dynamical evolution of 
$p_{m,n}(t)$, which determines $R(n,t)$ and all the
quantities of interest.

The systematic error in the results, due to the integration 
of Eq.~(\ref{eq2d}) up to a finite time, is fully negligible for realistic
values of $L$: 
the probability $Q(t)$ that a nucleation occurs at time
$t$ decays exponentially for large $t$ and consequently
the systematic error can easily be made exceedingly small. 
All numerical results presented in this paper can be considered
virtually exact.

As expected, the results for intermediate barriers
smoothly interpolate between the two limits of zero or infinite barriers,
being $\les/L$ the only relevant parameter.

\begin{figure}
\centerline{\psfig{figure=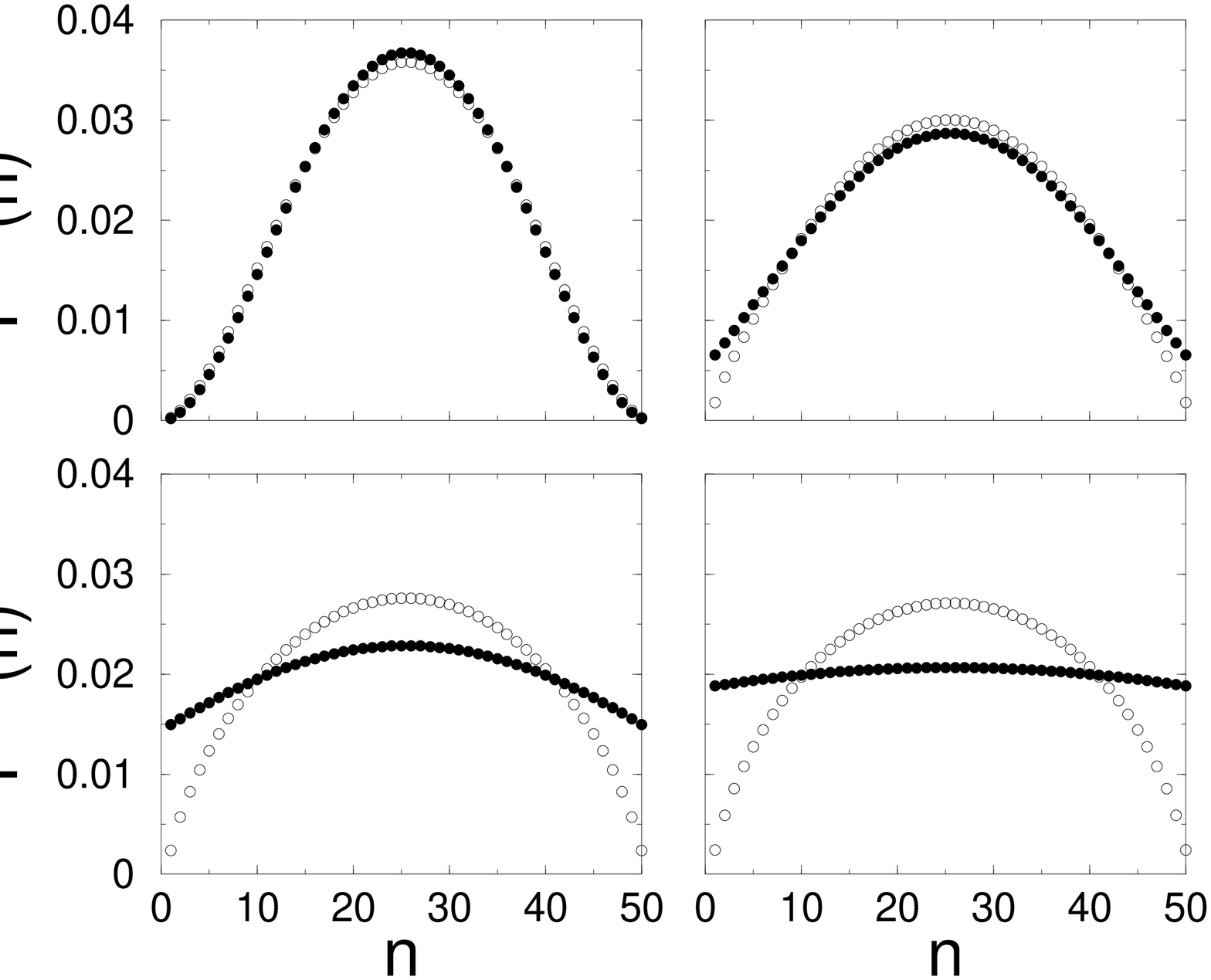,width=8cm,angle=-90}}
\caption{Normalized spatial distribution $P^{(N)}(n)$ for $d=1$ and
$L=50$. Empty circles are for interacting particles, full circles
for noninteracting particles (MFT).
$\les=0$ (top left), $\les=10$ (top right), $\les=50$ (bottom left),
$\les=250$ (bottom right).}
\label{fig5}
\end{figure}

The spatial distribution of nucleation events $P^{(N)}(n)$ is presented
in Fig.~\ref{fig5} for $L=50$ and several values of $\les$
($\les=0,10,50,250$).
Even a small value $\les/L=1/5$ changes in a notable way
the distribution $P^{(N)}(n)$.
The temporal distribution $Q(t)$ smoothly interpolates between
the two limit behaviors presented in Fig.~\ref{fig2}.

\section{Results in two dimensions}
\label{d2}
When the terrace is two-dimensional the motion of two adatoms
can be mapped into a four-dimensional problem for a single random walker:
$p_{m_1,n_1,m_2,n_2}(t)$ is the probability
of finding one atom on site $(m_1,n_1)$ and the other in $(m_2,n_2)$
at time $t$.
Such a probability obeys the equation of motion

\bea
\label{eq4d}
p_{m_1,n_1,m_2,n_2}(t+1)={1\over 8}\left[ \right.
& p_{m_1+1,n_1,m_2,n_2}(t)+p_{m_1-1,n_1,m_2,n_2}(t) + \\ \nonumber
& p_{m_1,n_1+1,m_2,n_2}(t)+p_{m_1,n_1-1,m_2,n_2}(t) + \\ \nonumber
& p_{m_1,n_1,m_2+1,n_2}(t)+p_{m_1,n_1,m_2-1,n_2}(t) + \\ \nonumber
& \left. p_{m_1,n_1,m_2,n_2+1}(t)+p_{m_1,n_1,m_2,n_2-1}(t) \right]
\eea
with the boundary condition $p_{\tilde n +\tilde\delta} =
a p_{\tilde n}$, where $\tilde n$ is any edge site of the four-dimensional
hypercube and $(\tilde n +\tilde\delta)$ is a nearest neighbour outside
the cube.
The initial condition is
\be
p_{m_1,n_1,m_2,n_2}(0)= \fra{1}{2}
[ \pu_{m_1,n_1} \ps_{m_2,n_2} + \ps_{m_1,n_1} \pu_{m_2,n_2} ] ~ ,
\label{ini4d}
\ee
where -- as usual -- $\pu_{m,n}= 1/L^2$ is the uniform initial distribution
in two dimensions and $\ps_{m,n}$ is the normalized stationary
solution of the discrete diffusion equation in $d=2$.

\subsection{Zero barriers (regime {\em i})}

\subsubsection{Noninteracting adatoms}

For noninteracting adatoms the computation of the quantities of interest
proceeds along the same lines as in the one-dimensional case.
The general solution of the equation of motion for the four-dimensional
random walker is
\bea
p_{m_1,n_1,m_2,n_2}(t)& = &\sum_{k_1,j_1,k_2,j_2=1}^L 
B_{k_1j_1k_2j_2} \\
& & {1\over 4^t}\left[
\cos\left({k_1\pi\over L+1}\right) 
+ \cos\left({j_1\pi\over L+1}\right)
+ \cos\left({k_2\pi\over L+1}\right)
+ \cos\left({j_2\pi\over L+1}\right)
\right]^t \\
& &
\sin\left({m_1 k_1\pi\over L+1}\right)
\sin\left({n_1 j_1\pi\over L+1}\right)
\sin\left({m_2 k_2\pi\over L+1}\right)
\sin\left({n_2 j_2\pi\over L+1}\right)
\eea
where the coefficients $B_{k_1j_1k_2j_2}$ are
\bea
B_{k_1j_1k_2j_2}& =& \left({2\over L+1}\right)^4 
\sum_{m_1,n_1,m_2,n_2=1}^L p_{m_1,n_1,m_2,n_2}(0) \\
& & 
\sin\left({m_1 k_1\pi\over L+1}\right)
\sin\left({n_1 j_1\pi\over L+1}\right)
\sin\left({m_2 k_2\pi\over L+1}\right)
\sin\left({n_2 j_2\pi\over L+1}\right) ~ .
\eea
Given the initial condition~(\ref{ini4d}), the coefficients
$B_{k_1j_1k_2j_2}$ are of the form
\be
B_{k_1j_1k_2j_2} = A_{k_1j_1}^U A_{k_2j_2}^S ~ ,
\ee
where $A_{kj}^{(U,S)}$ are the coefficients of the expansion of
$p_{m,n}^{(U,S)}$ (see Ref.~\cite{Politi01}).

The probability $R(m,n,t)$ of a (fictitious) nucleation event at time $t$ on
site $(m,n)$ is given by $p_{m,n,m,n}(t)$.

\begin{figure}
\centerline{\psfig{figure=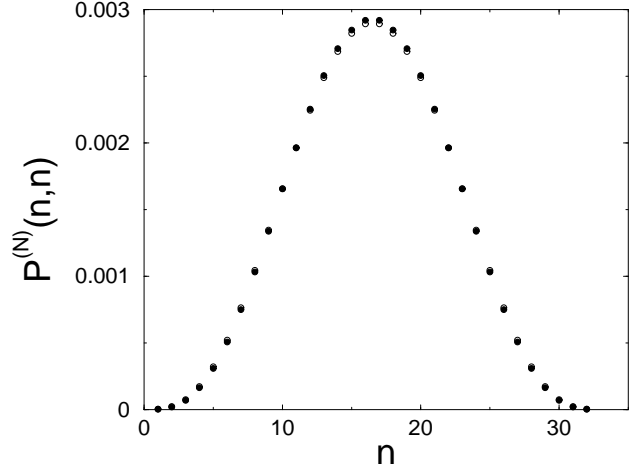,width=8cm,angle=-90}}
\caption{Normalized spatial distribution $P^{(N)}(n,n)$ along the diagonal
for $d=2$, $L=32$ and $\les=0$.
Empty circles are for interacting particles, full circles for
noninteracting particles.}
\label{fig6}
\end{figure}

\noindent $\bullet$ {\it Nucleation sites.}
The spatial distribution of nucleation sites is
\begin{equation}
\Pni(m,n) = \sum_{t=0}^\infty R(m,n,t) = \sum_{t=0}^\infty p_{m,n,m,n}(t)
\end{equation}
and its normalized version is reported in Fig.~\ref{fig6}
[we plot it along the diagonal of the square terrace, $\Pni^{(N)}(n,n)$].

\begin{figure}
\centerline{\psfig{figure=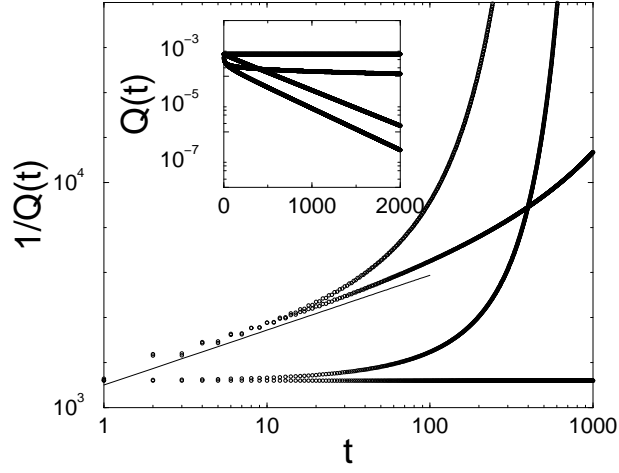,width=8cm,angle=-90}}
\caption{The temporal distribution for $d=2$ and $L=40$.
In the main part $1/Q(t)$ is plotted vs $t$ to highlight the logarithmic
decay for short times.
From top to bottom data are for noninteracting particles ($\les=\infty$
and $\les=0$) and interacting particles ($\les=\infty$ and $\les=0$).
The solid line goes as $\ln(t/t_0)$ [i. e. $Q(t)\sim 1/\ln(t/t_0)$].
The inset shows the exponential decay of $Q(t)$ for long times.}
\label{fig7}
\end{figure}

\noindent $\bullet$ {\it Nucleation times.}
The distribution of nucleation times is
\begin{eqnarray}
\Qni(t) & = & \sum_{m,n=1}^L p_{m,n,m,n}(t) \\
& = & \sum_{k_1,j_1,k_2,j_2=1}^L B_{k_1j_1k_2j_2} \\
&   & \left\{{1 \over 4} \left[
\cos\left({k_1 \pi \over L+1}\right)
+\cos\left({j_1 \pi \over L+1}\right) 
+\cos\left({k_2 \pi \over L+1}\right) 
+\cos\left({j_2 \pi \over L+1}\right) 
\right] \right\}^t \\
& & \sum_{m,n=1}^L
\sin\left({m k_1 \pi\over L+1}\right)
\sin\left({m k_2 \pi\over L+1}\right)
\sin\left({n j_1 \pi\over L+1}\right)
\sin\left({n j_2 \pi\over L+1}\right)
\\
& = & \left({L+1 \over 2}\right)^2
\sum_{k_1,j_1=1}^L B_{k_1j_1k_1j_1}
\left\{{1 \over 2} \left[
\cos\left({k_1 \pi \over L+1}\right)
+\cos\left({j_1 \pi \over L+1}\right) 
\right] \right\}^t
\end{eqnarray}
and it is plotted in Fig.~\ref{fig7}.
For large times, only the most slowly decaying mode $k_1=1, j_1=1$
contributes to the sum, yielding
\be
\Qni(t) \sim \exp \left[t \ln \cos \left({\pi \over L+1} \right)
\right] \simeq \exp \left[- {t \over 2} \left({\pi \over L+1} \right)^2 
\right] ~ .
\ee

\noindent $\bullet$ {\it Nucleation rate.}
The total number of times the two adatoms meet before leaving the terrace is
\begin{equation}
\Wni = \sum_{m,n=1}^L \Pni(m,n) = \left({L+1 \over 2}\right)^2
\sum_{k_1,j_1=1}^L
{B_{k_1j_1k_1j_1} \over 1-{1 \over 2} \left[
\cos\left({k_1 \pi \over L+1} \right)+
\cos\left({j_1 \pi \over L+1} \right)
\right]} ~ .
\end{equation}
For large values of $L$ only the mode $(1,1,1,1)$ dominates the sum, so:
\be
\Wni \simeq \left({L+1 \over 2}\right)^2
{B_{1111} \over 1-{1 \over 2} \left[
\cos\left({\pi \over L+1} \right)+
\cos\left({\pi \over L+1} \right)
\right]} \simeq {2 \over \pi^2} \left({L+1 \over L} \right)^6
\label{Wnid2}
\ee
and the total nucleation rate is
\be
\oni \simeq F L^2 {\tau_{res} \over \tau_{dep}}
\left({2 \over \pi^2}\right) \left({L+1 \over L} \right)^6.
\ee
Using the expression $\tres \simeq \fra{2^5}{\pi^6}{L^2 \over D}$ derived
in Ref.~\cite{Politi01}, the leading term in $L$ is
\be
\oni \simeq {64 \over \pi^8} {F^2 L^6 \over D}~.
\ee

\subsubsection{Interacting adatoms}

At odds with what happens for the one-dimensional case,
the trick of using an initial condition antisymmetric with respect
to particle interchange can not be used in two dimensions
for taking into account the interaction between particles.
The physical reason is that in two dimensions two particles
can swap their position without meeting.
As a consequence, the configuration space can not be split 
into two dynamically
disconnected regions, because the condition $p_{m,n,m,n}(t)=0$
holds on a two dimensional plane that does not divide the four-dimensional
configuration space into separate domains.
Hence it is not possible to implement the additional boundary condition
$p_{m,n,m,n}(t)=0$ by choosing the initial
condition to be antisymmetric.
We have not been able to overcome this problem analytically and therefore
for the interacting case we resort to the
numerical solution of Eq.~(\ref{eq4d}), which is easily performed,
and gives virtually exact results (see Sec.~\ref{intermediate}).
The results for $P^{(N)}(n,n)$ and $Q(t)$ are presented in
Figs.~\ref{fig6} and~\ref{fig7}, respectively.
The spatial distribution as given by MF theory agrees with exact results
even better than in $d=1$;
the short time decay for $Q(t)$ does not follow a power law but rather
a logarithmic one [$Q(t)\sim1/\ln(t/t_0)$].

For what concerns the total nucleation rate we find numerically
$W \simeq 0.25/\ln (L/1.3)$ and this implies
\be
\omega \simeq 0.008 {F^2 \over D} {L^6 \over \ln (L/1.3)} ~ .
\ee

\subsection{Strong and infinite barriers (regimes {\em ii} and {\em iii})}

\begin{figure}
\centerline{\psfig{figure=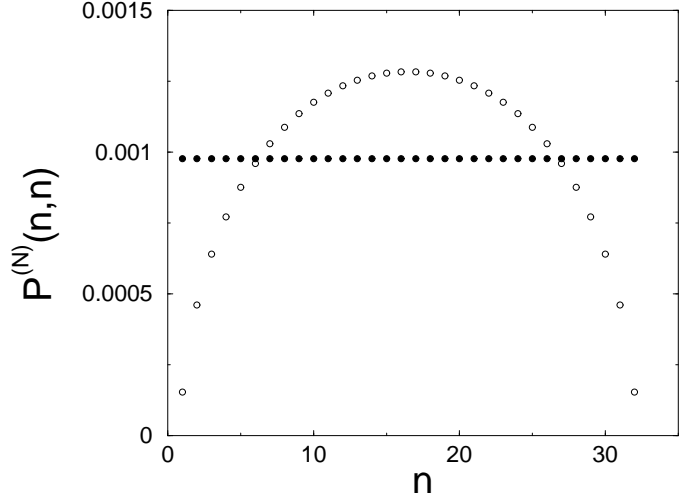,width=8cm,angle=-90}}
\caption{Normalized spatial distribution $P^{(N)}(n,n)$ along the diagonal
for $d=2$, $L=32$ and $\les=\infty$.
Empty circles are for interacting particles, full circles for
noninteracting particles.}
\label{fig8}
\end{figure}

\begin{figure}
\centerline{\psfig{figure=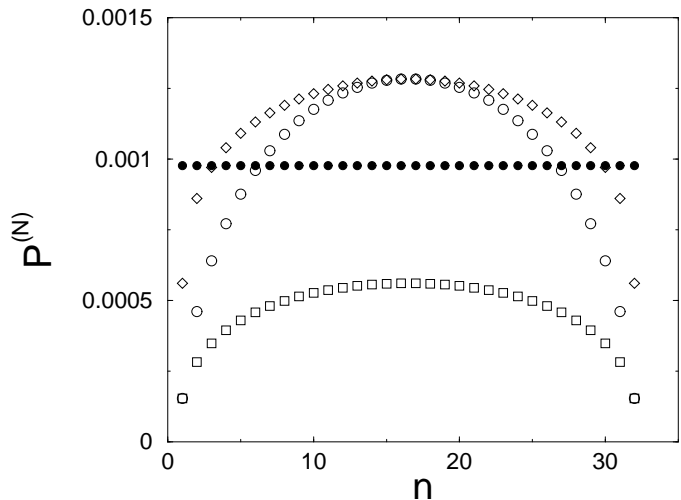,width=8cm,angle=-90}}
\caption{Normalized spatial distribution $P^{(N)}$ for $d=2$,
$L=32$ and $\les=0$.
Full circles are for noninteracting particles. Empty symbols are
for interacting particles: along the diagonal (circles), along one edge
(squares), and in the middle, parallel to one edge (diamonds).}
\label{fig9}
\end{figure}

The results for $Q(t)$ are presented in
Fig.~\ref{fig7} and those for $P^{(N)}(m,n)$ in 
Figs.~\ref{fig8},\ref{fig9}.
The spatial distribution along the diagonal (Fig.~\ref{fig8})
behaves much in the same way as in $d=1$;
a qualitatively similar behavior is found along
different directions (Fig.~\ref{fig9}). 
A deeper analysis is deferred to Sec.~\ref{disc_Pdin}.

The total number of nucleation events $W$ is clearly 1 and this implies,
in regime {\em ii}
\be
\omega = F L^2 {\tres \over \tdep} = {F^2 L^5 \les \over 4 D}
\label{omega2dles}
\ee
and in regime {\em iii}
\be
\omega = F L^2 = {1 \over \tdep} ~ .
\ee

\begin{figure}
\centerline{\psfig{figure=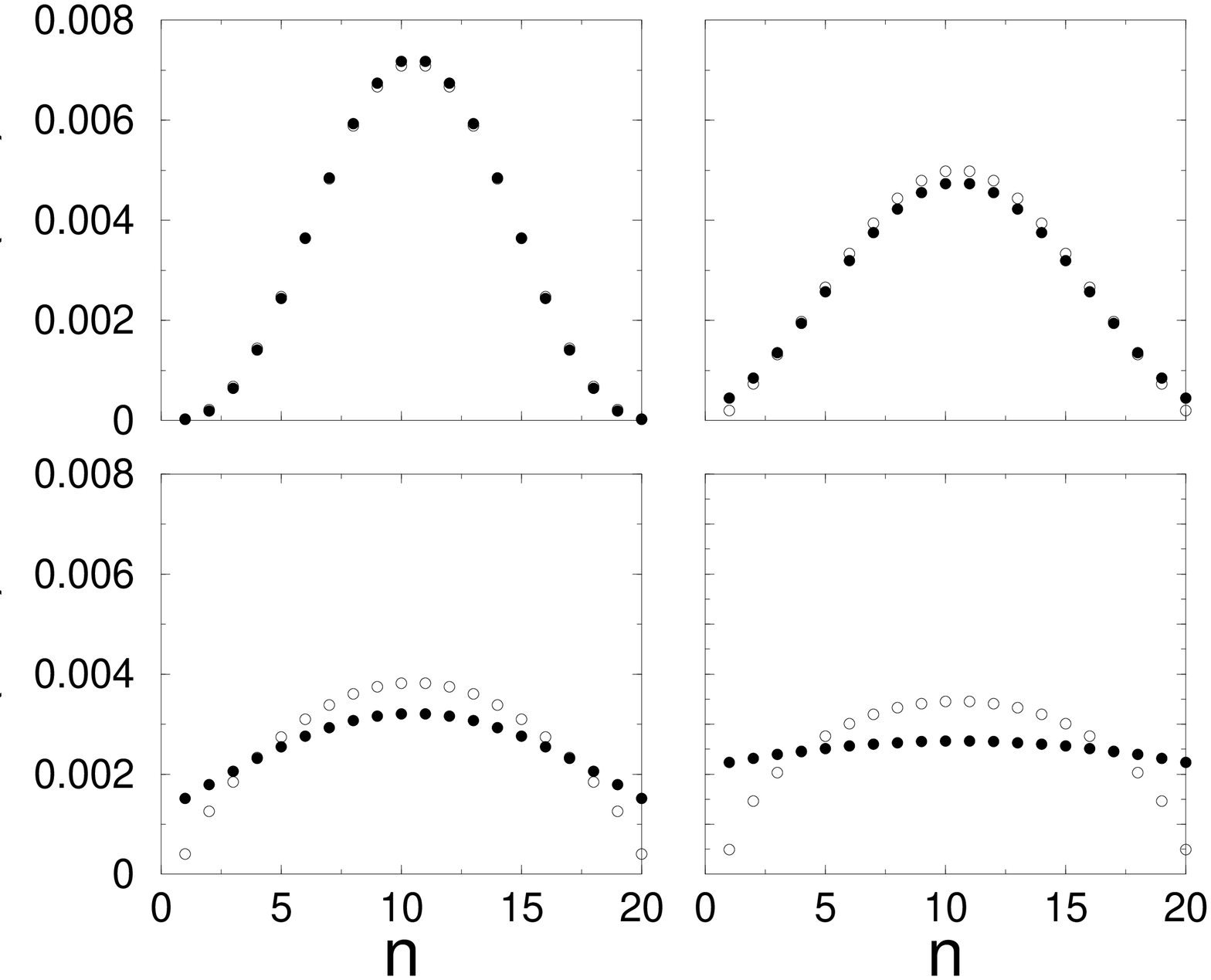,width=8cm,angle=-90}}
\caption{Normalized spatial distribution $P^{(N)}(n,n)$ along the
diagonal for $d=2$ and $L=20$.
Empty circles are for interacting particles, full circles
for noninteracting particles.
$\les=0$ (top left), $\les=4$ (top right), $\les=20$ (bottom left),
$\les=100$ (bottom right).}
\label{fig10}
\end{figure}

For intermediate barriers,
the results for $P^{(N)}(n,n)$ are presented in Fig.~\ref{fig10}.
As in $d=1$, the spatial distribution interpolates between the two
limits of zero and infinite barriers. As already remarked for
the one-dimensional case,
a relatively small ES barrier ($\les/L=1/5$) affects
quite dramatically the spatial distribution.

\section{Discussion of the results}
\label{disc}

\subsection{The spatial distribution}
\label{disc_Pdin}

The form of the spatial distribution of nucleation sites
has been presented in the preceding two Sections both in one and
two dimensions and for all values of the ES barrier.
Some remarks are in order.

As expected~\cite{Politi01}, we find that the mean-field assumption
for the distribution of nucleation sites {\em is in general not exact}
both in one and in two dimensions, for all values of $\les$ and for all $L$.
The origin of the discrepancy between the exact form of $P(n)$ and
the MF counterpart is clear: the mean-field approximation is
equivalent to considering particles as noninteracting, i. e., taking
into account not only the first nucleation event between the particles,
but also all subsequent encounters between them that would occur
should they keep diffusing after meeting.

Although not exactly the same, the mean-field distribution is however
{\em a very good approximation} of the true spatial distribution,
for zero or weak ES barriers, particularly in $d=2$ 
but also in $d=1$.
This result is somewhat striking,
if we consider that the ratio $\Wni/W$
is proportional to $L$ in $d=1$ [Eq.~(\ref{Wnid1})]
and to $\ln L$ in $d=2$ [Eq.~(\ref{Wnid2})].
Hence the relative weight of successive nucleations diverges for growing $L$;
nevertheless $\Pni(n)$ is very close to $P(n)$, indicating
that the distribution of all nucleation events following the first one is
very similar to the distribution of the first.

Things are radically different for large ES barriers.
In this case the discrepancy between MF and true distributions is remarkable.
Also this result is somewhat counterintuitive.
Particles are distributed uniformly at the beginning and each of them
would remain like that forever in the absence of the other:
this is the reason why the spatial distribution $\Pni$ for noninteracting
particles is uniform.
The interaction between particles breaks this uniformity. Consider for
example the one-dimensional case.
The nucleation probability on site $n$ close to the center of the
terrace, is the sum of the statistical weight of all pairs of random walks
(one for each particle) with the constraint that they intersect for
the first time in $n$.
If the site $n$ is close to an edge, one of these walks is
reflected by the boundary and the weight of walks intersecting for the
first time in $n$ is strongly reduced.

An `entropic' mechanism is present for weak barriers as well:
in this case nucleation close to an edge is made difficult by
adsorbing boundaries, which reduce the probability to find an atom close to
the steps. For weak barriers $P(n)$ is peaked around the middle of the
terrace also because the initial distribution for one atom is
not uniform, but parabolic.

Notice however that for infinite barriers
the mean-field distribution, which includes the
contribution of successive encounters, is completely flat. This indicates
that, even if it is relatively unusual for particles to meet close to edges,
once this happens they tend to meet there several times and
this restores uniformity in the distribution of all nucleation events.

In Eq.~(\ref{Papprox}) we have proposed an approximate expression for 
the distribution $P(n)$ in the limit of infinite ES barriers.
It has been derived assuming a behavior as an hyperbolic cosine
\be
P(n) = a_0 [a_1 - \cosh(a_2n -a_3)]
\ee
and imposing that $P(n)$ is symmetrical with respect the center 
of the terrace ($a_3/a_2=(L+1)/2$) and that
$P(0)=P(L+1)=0$ [$a_1=\cosh(a_3)$]. 
The former condition is obvious, and the latter derives from the
numerical evidence that $P(1)/P\left(\fra{L+1}{2}\right)$ goes to zero for
increasing $L$. Once both conditions have been imposed we obtain:
\be
P(n) = a_0 \left[\cosh a_3 - \cosh\left({2a_3n\over L+1} -a_3
\right) \right ]
\ee 
There is only one fit parameter, $a_3$, because $a_0$ 
is constrained by the normalization condition for $P(n)$. From
a nonlinear curve fitting for relatively small $L$
we can extrapolate that $a_3(L)$ tends to a constant value of order 3.11
as $L$ grows.
We have, somewhat arbitrarily, set $a_3(\infty)$ equal to $\pi$.

\subsection{The temporal distribution}
\label{disc_Qdit}

The results for the temporal distribution of nucleation events $Q(t)$
show in all cases a slow decay for short times
(as a power-law in $d=1$, logarithmic in $d=2$)
followed by an exponential decrease for larger times.
This behavior has been obtained by solving exactly, analytically
or numerically, the evolution equation for the particles on the terrace.
Its physical meaning is clarified further by rederiving these 
results by means of more transparent but less rigorous arguments:
the decay for short times is interpreted in terms of first passage properties
of random walks in an unbounded space;
the long time decay is the combined effect of the exponentially decreasing
probability that both particles are still on the terrace at time $t$
and the probability that they have not yet met.

Let us first discuss the behavior at short times and consider the
relative coordinate of the two particles as the coordinate of a
fictitious particle $C$: nucleation occurs when $C$ reaches the origin.
The initial spatial distribution probability for $C$
[$\rhoC(r)$] is a function of $\rho_A$ and $\rho_B$ complicated by
the presence of boundaries.
However, we are interested in the behavior for short times, i. e. 
times such that particles are not affected by the presence of terrace edges.
Therefore we can assume an initial spatial  distribution $\rhoC(r)$ 
uniform in a region of linear size $L$ around the origin ($\rhoC=1/L^d$)
and zero outside.
The irrelevance of boundaries in the short time regime is
confirmed by Figs.~\ref{fig2} and~\ref{fig7}: $Q(t)$ has the same behavior,
independently of step-edge barriers.

We now define $F(r,t)$ as the first passage probability
in $r$ at time $t$ starting from the origin at time zero.
The probability that atom $A$, leaving from $r$ at $t=0$ arrives for the
first time in the origin at time $t$ is clearly $F(-r,t)$, so that
\be
Q(t) = \sum_r \rhoC(r) F(-r,t) ~ .
\label{Q-F}
\ee
Let us also define $P(r,t)$ as the
probability that a particle is in $r$ at time $t$, being at the
origin at time zero. At $t=0$ we have $P(r,0)=\delta_{r,0}$ and
$F(r,0)=0$.
$P(r,t)$ and $F(r,t)$ are connected by~\cite{Hughes}
\be
P(r,t) = \sum_{\tau=0}^t F(r,\tau) P(0,t-\tau) ~ .
\label{eqP}
\ee
We write Eq.~(\ref{eqP}) for spatial argument $-r$, multiply
both sides by $\rhoC(r)$:
\be
\rhoC(r) P(-r,t) = \sum_{\tau=0}^t \rhoC(r) F(-r,\tau)P(0,t-\tau)
\ee
and sum over $r$:
\be
\sum_r \rhoC(r) P(-r,t) = \sum_{\tau=0}^t Q(\tau)P(0,t-\tau) ~ .
\ee

At short times $P(r,t)$ is negligible in the region where
$\rhoC(r)$ vanishes. Therefore we can take $\rhoC$ out of the summation
and use the normalization of $P(-r,t)$, obtaining:
\be
{1\over L^d} = \sum_{\tau=0}^t Q(\tau)P(0,t-\tau) ~ .
\label{eqP2}
\ee

In $d=1$, we pass to the continuuum in time 
[$P(0,t-\tau)=1/(t-\tau)^{1/2}$],
\be
{1\over L} = \int_0^t d\tau {Q(\tau)\over (t-\tau)^{1/2} }
\ee
and setting $\tau=t s $, we obtain
\be 
{1\over L} = t^{1/2} \int_0^1 ds {Q(ts)\over (1-s)^{1/2} } ~ ,
\ee
which implies $Q(t)\sim t^{-1/2}$.

In two dimensions we separate the term $\tau=t$ in
Eq.~(\ref{eqP2}):
\bea
{1\over L^2} &=& \int_0^{t-1} d\tau {Q(\tau)\over t-\tau} + Q(t)\\
       &=& \int_0^{1-1/t} ds {Q(ts) \over 1-s } + Q(t) \\
       &\simeq& Q(t) \int_0^{1-1/t} {ds \over 1-s } + Q(t) \\
       &\simeq& Q(t) [1+\ln t] 
\eea
and we obtain $Q(t)\sim 1/(1+\ln t)$.

In conclusion, at short times $Q(t)$ decays as a power law 
[$Q(t)\sim 1/\sqrt{t}$] in $d=1$ and logarithmically
[$Q(t)\sim 1/\ln(t/t_0)$] in $d=2$.

Let us consider now the behavior for long times.
The probability that a single adatom remains on the terrace up to time $t$
is (see Ref.~\cite{Politi01})
$ S(t) \sim \exp(-\alpha_S t)$.
In $d=1$ one has $\alpha_S = {1 \over 2} \left({\pi} \over L+1\right)^2$
for zero barriers and $\alpha_S=0$ for infinite barriers.
It is better to introduce a continuous time notation.
The time step for a single particle is $\Delta t=1/(2dD)$,
while for two particles diffusing on the same terrace is 
$\Delta t=1/(4dD)$, so that
\be
S(t) \sim \exp(-2\alpha_S dDt).
\label{SQ}
\ee

The probability that two adatoms meet at time $t$ decays as
\be
Q(t) \sim \exp(-4\alpha_Q dDt).
\ee
In Sec.~\ref{d1} we determined that for noninteracting particles 
$\alpha_Q={1 \over 2} \left({\pi} \over L+1\right)^2$ for zero barriers and
$\alpha_Q=0$ for infinite barriers, while
for interacting particles 
$\alpha_Q={5 \over 4} \left({\pi} \over L+1\right)^2$ for zero barriers and
$\alpha_Q={1 \over 4} \left({\pi} \over L\right)^2$ for infinite barriers.

All these findings are simply rationalized by the following argument.
We define $G(t)$ as the probability that two adatoms confined on the terrace
meet for the first time at time $t$: 
it is therefore equal to $Q(t)$ in the limit of infinite barriers.
For long times,
\be
G(t) \sim \exp(-4\alpha_G d Dt) ~ .
\ee

We claim that, for interacting adatoms, $Q(t)$ is given by the probability
that each of the two adatoms is still on the terrace times the probability
that they meet for the first time at time $t$:

\be
Q(t) \sim S^2(t) G(t)
\hspace{1cm}
\Rightarrow
\hspace{1cm}
\alpha_Q =  \alpha_S + \alpha_G ~ .
\label{alphaint}
\ee

In the noninteracting case, clearly $G(t)$ does not play any role. Then
\be
Q(t) \sim S^2(t)
\hspace{1cm}
\Rightarrow
\hspace{1cm}
\alpha_Q =  \alpha_S ~ .
\label{alphanon}
\ee

If we neglect the differences between $L$ and $L+1$ at the
denominators of $\alpha_Q$, the relations (\ref{alphaint}) and
(\ref{alphanon}) are both verified in the limits $\les=0$
and $\les=\infty$.

In $d=2$ the value of $\alpha_Q$ is not known analytically. However 
relations (\ref{alphaint},\ref{alphanon}) have been verified numerically.

\subsection{The nucleation rate}
\label{nuc_rate}

In this paper we have computed exactly the scaling of the
nucleation rate in all regimes, in $d=1$ and $d=2$, both
for the noninteracting (mean-field) and the interacting case.
The results confirm those of Ref.~\cite{Politi01},
where the rigorous calculation of $W$ was lacking.
Mean field theory overestimates the nucleation rate by a factor
that scales, in regime {\em i} (zero or weak barriers), as $L$ in $d=1$
and $\ln L$ in $d=2$.
In the limit of strong barriers (regime {\em ii}) the error scales as
$\les$ in $d=1$ and as $\les/L$ in $d=2$. Notice that the latter is
a large quantity, since in this regime $\les \gg L$.
For infinite barriers (regime {\em iii}) the mean-field picture
trivially breaks down.
Hence mean-field theory is generally strongly inaccurate, except in
two dimensions for weak barriers; however, even in this case
logarithmic corrections render $\omf$ not completely reliable.

Our treatment allows the evaluation not only of exponents, but
also of prefactors. In particular, this is performed analytically
for zero or strong barriers in $d=1$ [Eqs.~(\ref{omega1d})
and~(\ref{omega1dles})] and for strong barriers in $d=2$
[Eq.~(\ref{omega2dles})], while for $d=2$ and no barriers we have
evaluated the prefactor numerically.
For reference we report in Table~\ref{Table1} the value
of the nucleation rate in the different cases.

\begin{table}
\begin{center}
\begin{tabular}{|c|c|c|} 
& $d=1$ & $d=2$ \\ \hline
Weak barriers & $\omega \simeq 0.04 {F^2 L^4 \over D}$
& $\omega \simeq 0.008 {F^2 L^6 \over D \ln (L/1.3)}$ \\
(regime {\em i}) & & \\  \hline 
Strong barriers &$\omega = {1 \over 2} {F^2 L^3 \les \over D}$
& $\omega = {1 \over 4} {F^2 L^5 \les \over D}$\\
(regime {\em ii}) & & \\
\end{tabular}
\end{center}
\caption{Value of the nucleation rate $\omega$, including the
correct prefactors. In the regime {\it iii} of infinite
barriers, $\omega=(FL^d)^{-1}$. }
\label{Table1}
\end{table}

Finally, we want to remark that           
in $d=1$ for zero barriers, not only the asymptotic
behavior for large $L$, but the exact value of $\omega$ for any $L$, can
be determined analytically. One just needs to perform the sum of
$L^2$ terms [Eq.~(\ref{Wi})].

\section{Conclusions}
\label{concl}

In this paper and in the preceding one we have presented a rather
complete study of the problem of irreversible dimer nucleation on
top of terraces during epitaxial growth.
We have analyzed in detail the mean-field approach to this problem,
identified its weaknesses and provided a physical interpretation for
them.
Then we have solved the problem, by analytical means or
(when needed) numerically.
In this way we have derived exact results for the spatial and
temporal distributions of nucleation events and for the total nucleation
rate.

We believe that these results provide a relevant contribution to the
investigation of crystal growth both from the experimental and the
theoretical point of view.

The dependence of the nucleation rate $\omega$ on the terrace size $L$ and
the ES length $\les$ is a crucial piece of information for the
interpretation of experimental results, for example 
the evaluation of the Ehrlich-Schwoebel barrier.
The mean-field approximation has been widely used so far:
as already pointed out~\cite{Krug00}, this introduces a systematic 
underestimate of the strength of the ES barrier.
The exact expressions for $\omega$, derived in this work,
must replace the MF approximate formulas for a correct interpretation
of experimental data.

From the theoretical point of view, also the spatial distribution
plays an important role. Sometimes, `mesoscopic' models are used
to describe the growth process in the submonolayer regime~\cite{Ratsch}
or in the multilayer regime~\cite{EVPV}.
The rule for dimer formation must be supplemented with the spatial
distribution $P(n)$ of nucleation sites:
as we have argued, if additional step-edge barriers are not negligible, exact
results are completely different from mean-field predictions.

Let us finally mention some possible extensions of the present work.
In this paper and in the previous one, we have discussed 
irreversible nucleation on top of compact terraces:
it is therefore natural to wonder what occurs
if these hypotheses are relaxed.

The possibility of dimer dissociation introduces new time scales:
the average lifetimes of all unstable $j$-clusters
($2\le j\le i^*+1$).
Within the framework presented in this paper, this problem is mapped into
the random walk of a particle in a suitable high-dimensional space
with, in general, a spatially varying diffusion coefficient.
This inhomogeneity reflects the fact that unstable $j$-clusters
diffuse and break up with rates different from the single adatom
diffusion coefficient.
In many cases the full solution is therefore beyond reach,
even numerically (unless $i^*$ and $L$ are very small).

However in the simplest cases our approach may still be fruitful.
Let us consider for example $i^*=2$ and $d=1$:
three particles must meet (in the same lattice site) in order to
nucleate a stable trimer. When two particles meet they form a dimer
that dissociates after a typical time $\tdis$.
If $\tdis$ is much smaller than all other time scales,
two adatoms diffuse as they were noninteracting, the $3d$ walker
diffuses isotropically and we must just consider its
{\it irreversible} passage along the diagonal ($x_1=x_2=x_3$).
The same applies for generic $d$ and $i^*$ as
long as dissociation times of unstable clusters are small.
This case is also of interest to test recent scaling 
approaches~\cite{scaling} valid in the same limit ($\tdis\to 0$).

The second natural extension of the present work consists in considering
nucleation on top of fractal islands instead of compact ones. 
The framework of our method keeps unchanged~\cite{art_prepa}.

A further extension is to take into account the possibility
of re-evaporation of deposited particles.

\appendix

\section{Asymptotic behaviors of the temporal distribution in $d=1$.}
\label{App2}
In this appendix we present some detailed results for the one-dimensional
case with zero ES barriers and both adatoms having a uniform
initial distribution.
This is of course not physically sensible, since the effective
initial distribution of the second adatom has a parabolic form
for $\les=0$. However, contrary to the physically sensible case,
the evaluation of the coefficients $B_{kj}^<$ is not difficult
and this allows an explicit analytic evaluation of the behavior of
$Q(t)$, which will differ from the realistic one only in the prefactors.

A simple, although lengthy, evaluation of the coefficients leads to
\be
B_{kj}^< 
= {2\over (L+1)^2 L^2}\left[ \cot\left({1\over 2}{k\pi\over L+1}\right)
\cdot S_1 -\csc\left({1\over 2}{k\pi\over L+1}\right) \cdot S_2 \right]~,
\ee
where
\be
S_1 = \sin\left({L\over 2}{j\pi\over L+1}\right)
\sin\left({j\pi\over 2}\right)
\csc\left({1\over 2}{j\pi\over L+1}\right)
\ee
and
\bea
S_2 &=& {1\over 2}
\sin\left[-{1\over 2}{k\pi\over L+1}+{(j+k)\pi \over 2}\right]
\sin\left[{L(j+k)\pi\over 2(L+1)}\right]
\csc\left[{j+k\over 2}{\pi\over L+1}\right] \\
&& +{1\over 2}
\sin\left[{1\over 2}{k\pi\over L+1}+{(j-k)\pi \over 2}\right]
\sin\left[{L(j-k)\pi\over 2(L+1)}\right]
\csc\left[{j-k\over 2}{\pi\over L+1}\right]~.
\nonumber
\eea

We now want to calculate the temporal distribution
\begin{eqnarray}
Q(t+1>0) & = & \sum_{n=1}^L {\tilde p}_{n-1,n}(t) \\
& = & \sum_{k,j=1}^L B_{kj} C_{kj} \left\{ {1 \over 2} \left[
\cos\left({k \pi \over L+1} \right) +
\cos\left({j \pi \over L+1} \right) \right] \right\}^t
\end{eqnarray}
where the coefficients $C_{kj}$ are given in Eq.~(\ref{C}).

In the limit of large $L$, the coefficients $C_{kj}$ are nonvanishing
only for odd $j-k$ (except for $k=j$, but $B_{kk}=0$) and their value
is
\be
C_{kj} = -k\left({1\over j-k} + {1\over j+k}\right) = {2kj\over k^2-j^2}~.
\ee
In the same limit
\be
S_1  = {2L\over j\pi} \sin^2 {j\pi\over 2}~,
\ee
\be
S_2 = {L\over \pi} {2j\over j^2-k^2}~.
\ee

Therefore:
\be
B_{kj}^< = {8\over L^2 k\pi^2} \left[
{1\over j}\sin^2{j\pi\over 2} - {j\over j^2-k^2} \right]~,
\ee
and after some algebra
\be
B_{kj} = {8\over\pi^2 L^2} {1\over jk}\left[
(-1)^k + {k^2+j^2\over k^2-j^2}\right]
\ee
and
\be
B_{kj}C_{kj} = {16\over\pi^2 L^2}\left[
{(-1)^k\over k^2-j^2} + {k^2+j^2\over (k^2-j^2)^2}\right]~.
\ee
The dominant contribution to $Q(t)$ comes from $j=k-1$ or $j=k+1$
\be
B_{k,k\pm 1}C_{k,k\pm 1} = {16\over\pi^2 L^2}\left[
{(-1)^k\over \mp 2k-1} + {2k^2\pm 2k +1 \over (\mp 2k-1)^2}\right]~.
\ee
Hence
\bea
Q(t) &\simeq {16 \over \pi^2 L^2 } & \left\{  \sum_{k=1}^{L-1}
\left[{(-1)^k\over -2k-1} + {2k^2 + 2k +1 \over (-2k-1)^2} \right]
\exp \left[-{t \over 4} \left({\pi \over L+1}\right)^2 (2k^2+2k+1)
\right] \right. \\
& & \left. + \sum_{k=2}^{L}
\left[{(-1)^k\over 2k-1} + {2k^2 - 2k +1 \over (2k-1)^2} \right]
\exp \left[-{t \over 4} \left({\pi \over L+1}\right)^2 (2k^2-2k+1)
\right] \right\}
\eea
We neglect the first term in both sums
and approximate the sums according to the Euler-Maclaurin formula
\be
\sum_{k=1}^L f_k = \int_{1}^L dk f(k) + {1\over 2}[f_1 + f_L]
\ee
Since in this case $f_L\ll f_1$, we neglect $f_L$:
\bea
Q(t) &\simeq {16 \over \pi^2 L^2 } & \left\{  \int_1^{L-1} dk
{2k^2 + 2k +1 \over (-2k-1)^2}
\exp \left[-{t \over 4} \left({\pi \over L+1}\right)^2 (2k^2+2k+1)
\right] \right. \\
& & + {5 \over 9} \exp \left[-{5 t \over 4} \left({\pi \over L+1}\right)^2
\right] \\
& & \left. + \int_2^L dk
{2k^2 - 2k +1 \over (2k-1)^2}
\exp \left[-{t \over 4} \left({\pi \over L+1}\right)^2 (2k^2-2k+1)
\right] \right\}~.
\eea

The integrals can be evaluated by considering for the integrand
the limit for large $k$, and one obtains:
\bea
Q(t) &\simeq {8 \over \pi^2 L^2 } {L \over \pi \sqrt{t/2}} 
\left[ \int_{ \pi\sqrt{t/2}/L}^{\pi\sqrt{t/2}} dx e^{-x^2}
      + \int_{2\pi\sqrt{t/2}/L}^{\pi\sqrt{t/2}} dx e^{-x^2} \right] \\
& + {80 \over 9 \pi^2 L^2}
\exp \left[-{5 t \over 4} \left({\pi \over L+1}\right)^2 \right]~.
\eea
The upper limit of the integrals can be shifted to $\infty$
\bea
Q(t) &\simeq {8 \over \pi^3 L \sqrt{t/2}} 
\left[ \int_{ \pi\sqrt{t/2}/L}^{\infty} dx e^{-x^2}
      + \int_{2\pi\sqrt{t/2}/L}^{\infty} dx e^{-x^2} \right] \\
& + {80 \over 9 \pi^2 L^2}
\exp \left[-{5 t \over 4} \left({\pi \over L+1}\right)^2 \right]~.
\eea

If $t\ll (2 L^2/\pi^2)$ each integral goes to $\sqrt{\pi}/2$ and the first
term prevails on the exponential; in the limit $t\gg (2 L^2/\pi^2)$
the opposite is true. So, $Q(t)\simeq 8/(L \pi^{5/2} \sqrt{t/2})$ 
for $t\ll (2 L^2/\pi^2)$
and $Q(t)\simeq {80 \over 9 \pi^2 L^2} \exp
\left[-{5 t \over 4} \left({\pi \over L+1} \right)^2 \right]$
for $t\gg (2 L^2/\pi^2)$.

By summing $Q(t)$ over $t$ one obtains $W$
\be
W  \simeq  \int_0^{2 L^2/\pi^2} dt Q(t) = 
\int_0^{2 L^2/\pi^2} dt {8 \over L \pi^{5/2} \sqrt{t/2}}
 \simeq  {32 \over \pi^{7/2}} \simeq 0.58\ldots
\ee

\end{document}